\pdfoutput=1
\documentclass[journal]{IEEEtran}

\usepackage{amsmath,amssymb,amsfonts}
\usepackage{graphicx}
\usepackage{cite}
\usepackage{booktabs}
\usepackage{multirow}
\usepackage{url}
\usepackage{xspace}
\usepackage{dblfloatfix, subcaption}
\usepackage[hidelinks]{hyperref}

\newcommand{\tqir}{\textsc{TQIR}\xspace}
\newcommand{\qtip}{\textsc{QTIP}\xspace}
\newcommand{\SWA}[3]{$S_{#1}W_{#2}A_{#3}$}
\begin{document}

\title{TQIR: Trellis Quantization with Integer Reconstruction for Bandwidth-Efficient Edge LLM Acceleration}

\author{Kuang-Ting Huang and Tian-Sheuan Chang,~\IEEEmembership{Senior Member,~IEEE}%
\thanks{This work was supported by the National Science and Technology Council, Taiwan, under Grants 111-2622-8-A49-018-SB, 110-2221-E-A49-148-MY3, 113-2221-E-A49-078-MY3, and 113-2640-E-A49-005. The authors are with the Institute of Electronics, National Yang Ming Chiao Tung University, Hsinchu 30010, Taiwan (e-mail: a0922256995@gmail.com; tschang@nycu.edu.tw).}%
}

\maketitle

\begin{abstract}

Autoregressive large language model (LLM) inference is often limited by external-memory bandwidth, while existing ultra-low-bit trellis quantization reconstructs floating-point weights and leaves runtime Hadamard-transform and decoder costs unresolved. This paper presents TQIR, an algorithm--hardware co-design that decouples storage precision ($S$) from decoded-weight ($W$) and activation ($A$) compute precisions, enabling a better accuracy--hardware trade-off. TQIR performs trellis matching with floating-point candidates offline while constraining each candidate to an INT8 or INT4 codeword that can be reproduced exactly for integer GEMM at runtime. To reduce transform cost while preserving the integer weight path, TQIR relocates the reverse Hadamard transform to the activation side and employs an $H_{128}$ Subspace Hadamard Transform, together with a subspace-aligned $128\times2$ trellis tile that preserves the original 256-weight sequence length. The configurable decoder, synthesized in a TSMC 16-nm process, delivers 307.2~Gweights/s in 0.0509~mm$^2$ and achieves 13.93$\times$ and 1.52$\times$ the normalized decoder-area efficiency of Ecco and Codeword Decomposition, respectively. These results show that fixed-rate trellis compression can reduce weight traffic while sustaining an integer-GEMM datapath matched to the evaluated memory and PE rates.

\end{abstract}


\section{Introduction}
Large language models (LLMs) have enabled increasingly capable language understanding, reasoning, and generation, but their growing parameter counts impose substantial storage, memory-transfer, and computation costs. These costs are particularly restrictive for edge inference, where external-memory bandwidth, power, and silicon area are limited. On-device execution nevertheless remains attractive because it reduces communication latency, supports offline operation, and keeps sensitive data local. Efficient edge deployment therefore requires a representation that reduces the number of weight bits fetched for every token without replacing the original bandwidth bottleneck with expensive decompression or floating-point preprocessing.

The problem is most acute during batch-one autoregressive decoding. Each generated token invokes matrix--vector operations with limited weight reuse, so most model weights must be streamed again from external memory. Conventional scalar and group-wise quantization reduce this traffic, while outlier-smoothing and rotation improve low-bit robustness~\cite{expresult_actsmooth_smoothquant,expresult_methodactsmooth_quarot,quip_incoherence}. However, aggressive reduction of the stored weight width also constrains the runtime arithmetic format and can cause model-quality degradation. Entropy-coded schemes provide additional compression, but variable-length parsing complicates high-throughput decoding; Ecco, for example, combines nonuniform quantization with Huffman coding~\cite{expresult_entropybasedmethod_ecco}.

QTIP~\cite{qtip} provides an attractive alternative through high-dimensional trellis-coded quantization (TCQ), incoherence processing, a fixed-rate bitshift trellis, and computed Gaussian codes. Its local bit windows support parallel reconstruction without a large explicit codebook or sequential entropy decoding. However, the original inference path reconstructs FP16 weights, requires runtime Hadamard processing, and does not provide a bandwidth-matched hardware decoder. These limitations motivate jointly considering the numerical representation, transform placement, and decompression architecture.

This work presents TQIR, an algorithm--hardware co-design that converts QTIP's ultra-low-bit representation into a PE-ready integer stream by jointly addressing integer-reproducible trellis encoding, runtime Hadamard processing, and bandwidth-matched decompression. The main contributions are as follows.

\begin{itemize}
\item \textit{FP-Encoded INT-Decoding} performs accuracy-sensitive trellis matching in floating point during offline encoding, but constrains the candidates to values reproducible as INT8 or INT4 codewords. Online reconstruction emits an integer weight and a format-specific shared scale, eliminating per-weight FP16 reconstruction before integer GEMM.
\item TQIR relocates the reverse Hadamard transform to the activation path to preserve the integer weight path and smooth activation outliers before quantization. An \textit{$H_{128}$ Subspace Hadamard Transform (SHT)} reduces the total runtime Hadamard operations of LLaMA-2-7B by 49.5\% while retaining model quality close to the global transform.
\item \textit{Subspace-Aligned Trellis Encoding} identifies an interaction between subspace rotation and TCQ layout. Reshaping the 256-weight encoding tile to follow each Hadamard-subspace boundary improves perplexity without increasing trellis dimension, bit rate, or online decoding complexity.
\item A bandwidth-matched hardware implementation provides 256 decoded weights/cycle. A configurable decoder supports three storage/compute modes with one datapath, whereas format-customized decoders exploit overlap between adjacent bitshift windows using cross-input common-subexpression elimination (CSE). A pair-folded SHT architecture and layer-level schedule can hide the activation-transform latency under the evaluated output-stationary layer schedule.
\end{itemize}

The remainder of this paper presents the TQIR algorithm in Section~II, its hardware architecture in Section~III, the experimental evaluation in Section~IV, and conclusions in Section~V.

\section{TQIR: Algorithm--Hardware Co-Design}
\subsection{TQIR Overview and Decoupled Precision}

\begin{figure*}[t]
    \centering
    \includegraphics[width=0.7\textwidth]{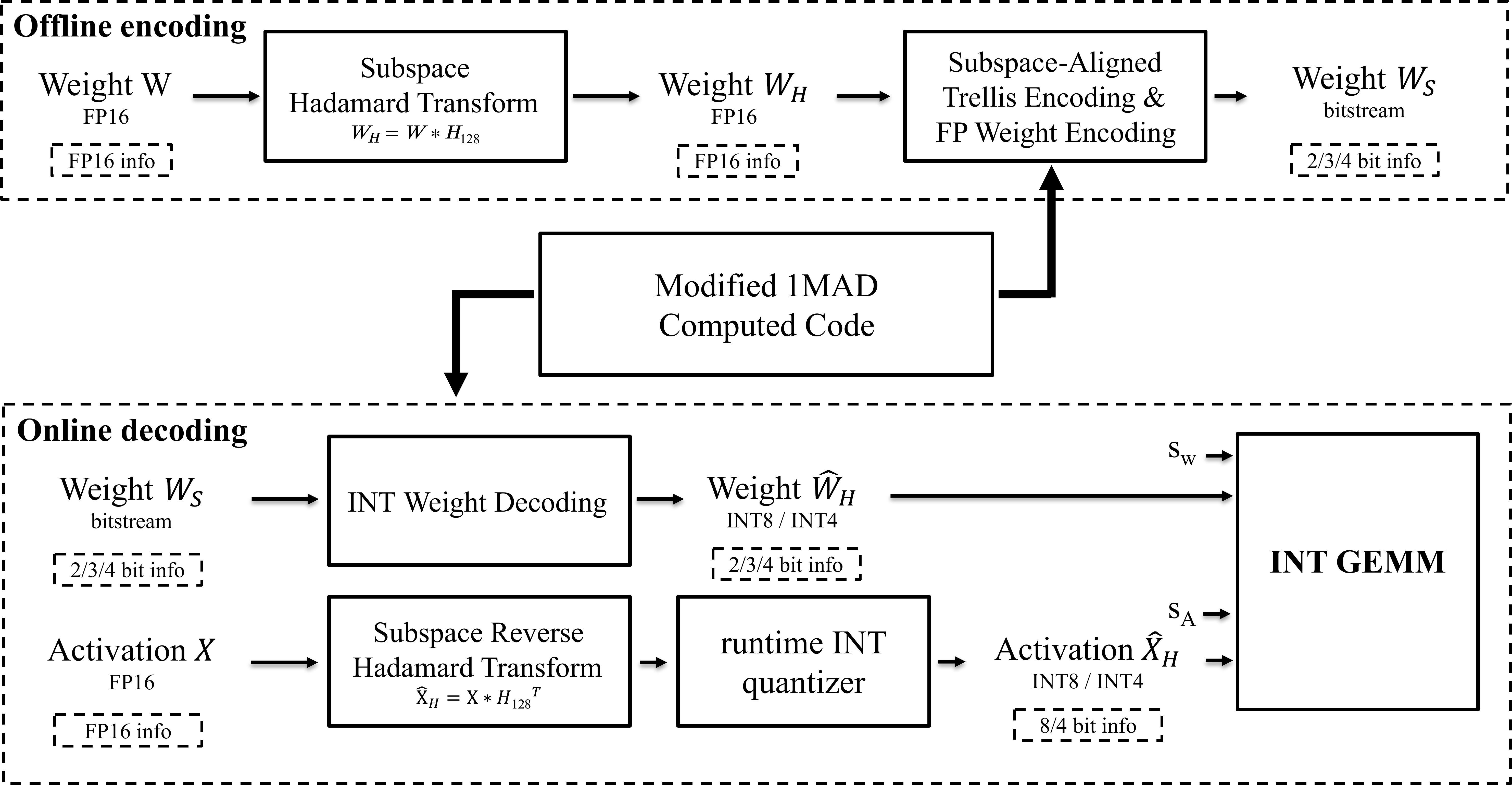}
    \caption{TQIR overview. Offline encoding performs subspace incoherence processing, subspace-aligned trellis search, and floating-point distortion matching over integer-reconstructable codewords. During online inference, the fixed-rate stream is decoded directly into INT8/INT4 weights, while the activation-side subspace transform precedes activation quantization and integer GEMM.}
    \label{fig:tqir_overview}
\end{figure*}

Fig.~\ref{fig:tqir_overview} summarizes the TQIR flow. Offline, FP16 weights are transformed within $B$-element subspaces and grouped into subspace-aligned 256-weight TCQ sequences. For each trellis state, the modified 1MAD mapping generates a candidate constrained to the target INT8 or INT4 grid and rescales it to floating point for distortion evaluation. Viterbi search then selects the minimum-distortion path, while only the 2-, 3-, or 4-bit trellis transitions are stored.

Online, overlapping state windows are decoded directly into INT8 or INT4 weights, while the activation-side SHT precedes activation quantization and integer GEMM; the associated scales are folded into subsequent requantization. We denote each mode as \SWA{x}{y}{z}, where $S_x$ is the stored trellis width, $W_y$ the decoded-weight format, and $A_z$ the activation format. The evaluated modes are \SWA{4}{8}{8}, \SWA{3}{4}{4}, and \SWA{2}{4}{4}. Although the hardware supports all three modes, changing the decoded-weight precision generally requires loading the corresponding offline-reencoded trellis stream.

\subsection{FP-Encoded INT-Decoding}
\label{subsec:fp_int_decoding}
\subsubsection{Formulation}

QTIP's 1MAD code maps a local $L$-bit trellis window $x$ through a linear congruential generator (LCG), sums its four output bytes, and subtracts the mean. With fixed unsigned 32-bit constants $a=34{,}038{,}481$ and $b=76{,}625{,}530$, the centered integer is
\begin{equation}
\begin{aligned}
p(x)&=(ax+b)\bmod 2^{32},\\
u(x)&=\sum_{i=0}^{3}\operatorname{byte}_i(p(x))-510,
\end{aligned}
\label{eq:one_mad}
\end{equation}
where $u(x)\in[-510,510]$ and its standard deviation is $\sigma_u\approx147.8$. QTIP uses the floating-point candidate $u(x)/\sigma_u$. TQIR instead incorporates the target integer grid before Viterbi search. For decoded format $b\in\{\mathrm{INT8},\mathrm{INT4}\}$, the format shift is $d_b=2$ or 6, and 
\begin{equation}
q_b(x)=\operatorname{clip}\!\left(\left\lfloor\frac{u(x)}{2^{d_b}}\right\rfloor,Q_{\min}^{b},Q_{\max}^{b}\right),\qquad c_b=\frac{2^{d_b}}{\sigma_u}.
\label{eq:int_grid}
\end{equation}
$\mathcal{Q}^b$ denotes the symmetric representable range: $[-127,127]$ for INT8 and $[-7,7]$ for INT4. 
The offline trellis candidate is $\widehat{w}_{H,b}(x)=c_bq_b(x)$. Ignoring clipping and flooring, the power-of-two contraction and restoration cancel, so $\widehat{w}_{H,b}(x)\approx u(x)/\sigma_u$. Crucially, the Viterbi search now evaluates candidates that can be reproduced exactly from $q_b(x)$ and the shared format constant $c_b$.

\begin{figure}[htbp]
    \centering
    \begin{subfigure}{\linewidth}
        \centering
        \includegraphics[width=0.9\linewidth]{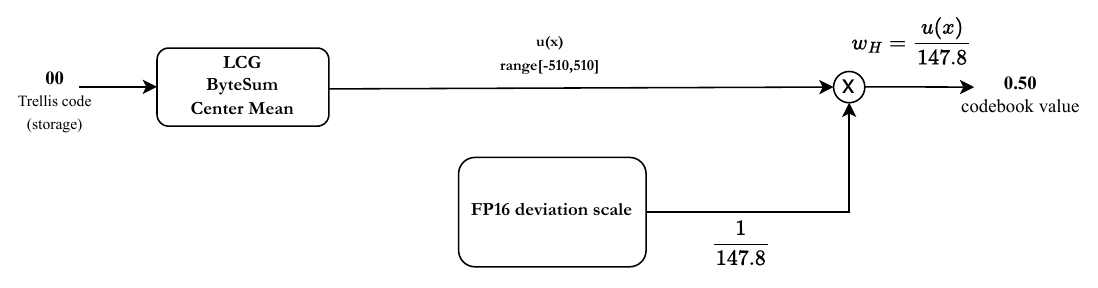}
        \caption{QTIP's 1MAD encoding procedure}
        \label{fig:QTIP_encoding_procedure}
    \end{subfigure}
    \begin{subfigure}{\linewidth}
        \centering
        \includegraphics[width=0.9\linewidth]{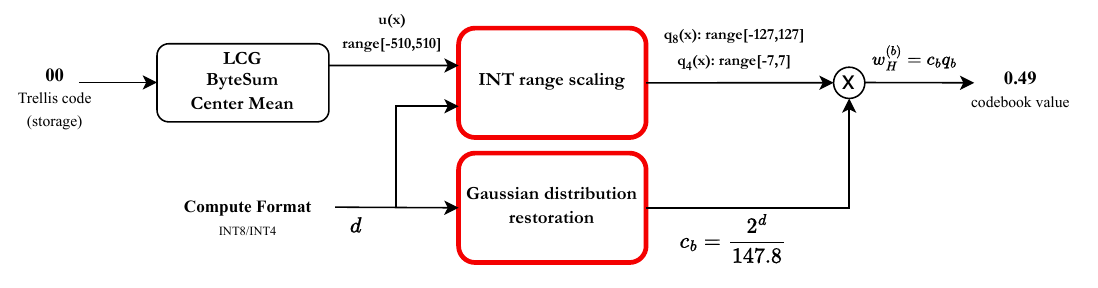}
        \caption{TQIR's floating-point weight encoding procedure}
        \label{fig:TQIR_FP_encoding_procedure}
    \end{subfigure}
    \caption{Floating-point weight-encoding procedure for FP-Encoded INT-Decoding. QTIP directly normalizes the centered 1MAD integer into a floating-point codeword, whereas TQIR first maps the value onto the target integer grid through INT Range Scaling and then restores the Gaussian scale for FP-domain trellis matching.}
    \label{fig:FP_encoding_procedure}
\end{figure}

During inference, the decoder returns $W_{\mathrm{INT}}=q_b(x)$ and forwards the scale factor $s_W=c_b$ as side information rather than multiplying every weight by $c_b$. This factor also motivates the use of 1MAD instead of the alternative 3INST computed code. Specifically, 1MAD maps a bounded integer to INT8/INT4 through fixed shifts, whereas 3INST constructs values through FP16 exponent and mantissa manipulation. For an activation quantizer producing the quantized value $A_{\mathrm{INT}}$ and the scale factor $s_A$, the linear output is
\begin{equation}
Y=s_Ws_A\left(W_{\mathrm{INT}}A_{\mathrm{INT}}\right).
\label{eq:scaled_int_gemm}
\end{equation}
The GEMM therefore uses integer multiplication and accumulation; $s_Ws_A$ is folded into output scaling. Relative to post-decoding quantization, the proposed method avoids selecting a floating-point path that may map poorly onto the target integer grid. Relative to QTIP, it removes per-weight normalization and FP16 reconstruction from the online weight path. The above formulation can be reduced to the original QTIP FP16 formulation by setting $d_b=0$ and $\mathcal{Q}_b = [-510,510]$. Fig.~\ref{fig:FP_encoding_procedure} shows the difference between the original QTIP flow and the proposed TQIR flow. 

\subsection{Activation-Side Reverse Hadamard and Subspace Transform}
FP-Encoded INT-Decoding enables INT8 or INT4 weights, thereby eliminating floating-point reconstruction from the online weight-decoding path. However, because the weights are encoded in the Hadamard-transformed domain, a reverse transform is needed at runtime to preserve the original linear operation. TQIR adopts activation-side reverse Hadamard transformation as shown in Fig.~\ref{fig:tqir_overview}, which is often used by rotation-based quantization methods such as QuaRot~\cite{expresult_methodactsmooth_quarot}, but here it serves an additional system-level purpose: keeping the decoded-weight path entirely in the integer domain while the activation transform is performed before activation quantization.

Even after relocating the Hadamard transform, a global transform remains expensive for long hidden dimensions. We therefore partition the dimension into independent $B$-element subspaces using $H'_B=I_{N/B}\otimes H_B$. A global $N$-point fast Walsh–Hadamard transform (FWHT) requires $N\log_2N$ element operations, whereas the subspace implementation requires only $N\log_2B$. The resulting theoretical reduction is
\begin{equation}
R_{\rm save}(N,B)=1-\frac{\log_2B}{\log_2N}.
\label{eq:shtsaving}
\end{equation}
With $B=128$, the projection-weighted transform count for LLaMA-2-7B decreases from 540,672 to 273,280 butterfly pairs per Transformer block, a reduction of 49.5\%. As shown in Table~\ref{tab:hadamard_cost}, both attention and feed-forward projections contribute substantially to this saving.

\begin{table}[t]
\centering
\caption{Runtime Hadamard butterfly-pair cost per projection on LLaMA-2-7B.}
\label{tab:hadamard_cost}
\scriptsize
\setlength{\tabcolsep}{3.2pt}
\begin{tabular}{lrrr}
\toprule
Projection group & Global & $H_{128}$ & Saving \\
\midrule
Q/K/V/O ($\times4$) & 49,152 & 28,672 & 41.6\% \\
Up/Gate/Down ($\times3$) & 114,688 & 52,864 & 53.9\% \\
\midrule
Total / Transformer block & 540,672 & 273,280 & \textbf{49.5\%} \\
\bottomrule
\end{tabular}
\end{table}

\subsubsection{Subspace-Aligned Trellis Encoding}

\begin{figure}
    \centering
    \includegraphics[width=0.7\columnwidth]{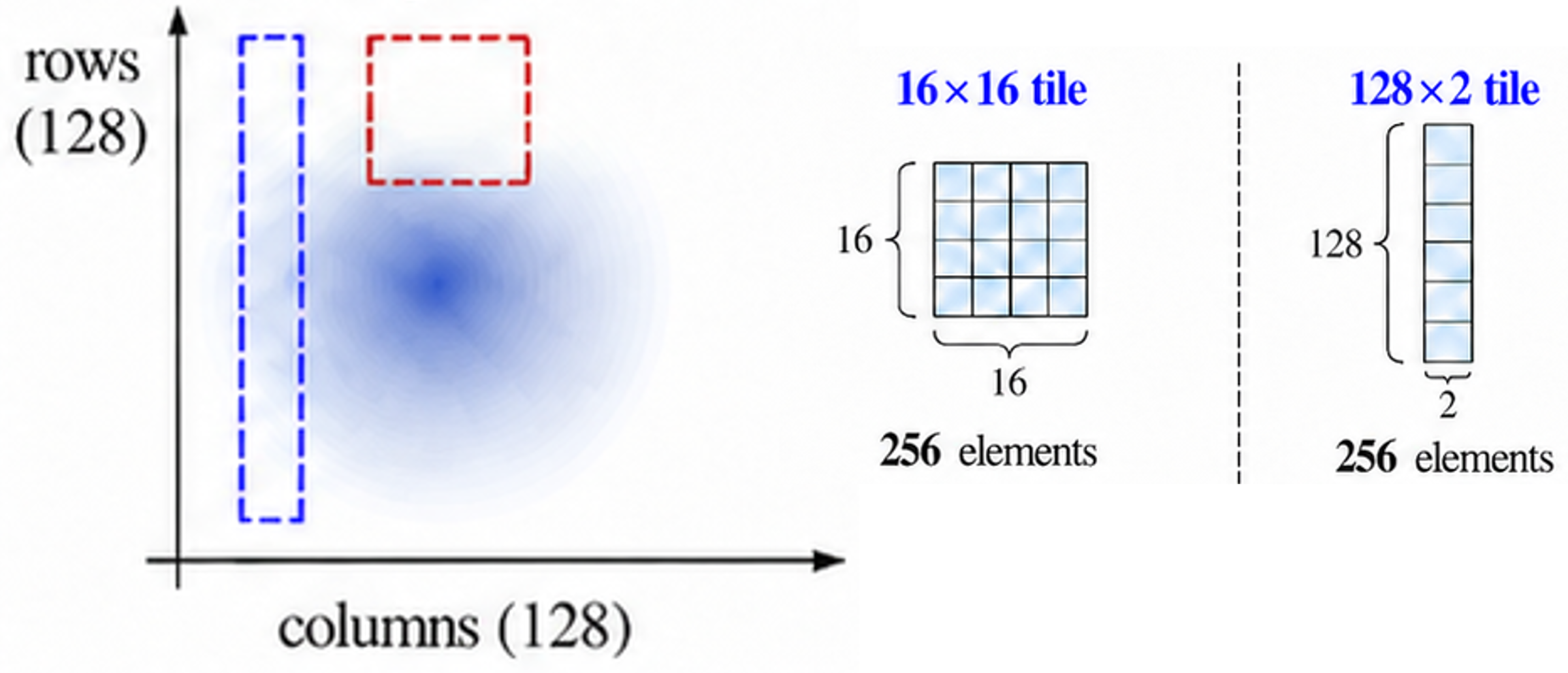}
    \caption{Subspace-aligned trellis encoding}
    \label{fig:subspace}
\end{figure}

The original \qtip implementation organizes each 256-weight trellis sequence as a $16\times16$ tile. After the Hadamard transform is partitioned into independent subspaces, however, this square tile covers only a localized portion of a larger subspace, which can reduce statistical coverage within each subspace and cause measurable perplexity degradation. Fig.~\ref{fig:subspace} illustrates the coverage of different tile shapes over the Gaussian distribution. \tqir therefore reshapes the same 256 weights so that the trellis sequence extends along the full subspace dimension:
\begin{equation}
\mathcal{T}_B\in\mathbb{R}^{B\times(256/B)},\qquad |\mathcal{T}_B|=256.
\label{eq:tile}
\end{equation}
Accordingly, $H_{128}$ uses a $128\times2$ tile while preserving the state length, trellis dimension, storage bit rate, and online decoder.

\section{Bandwidth-Matched Hardware Architecture}

\subsection{System-Level Integration}
\begin{figure}[htbp]
    \centering
    \includegraphics[width=0.8\linewidth]{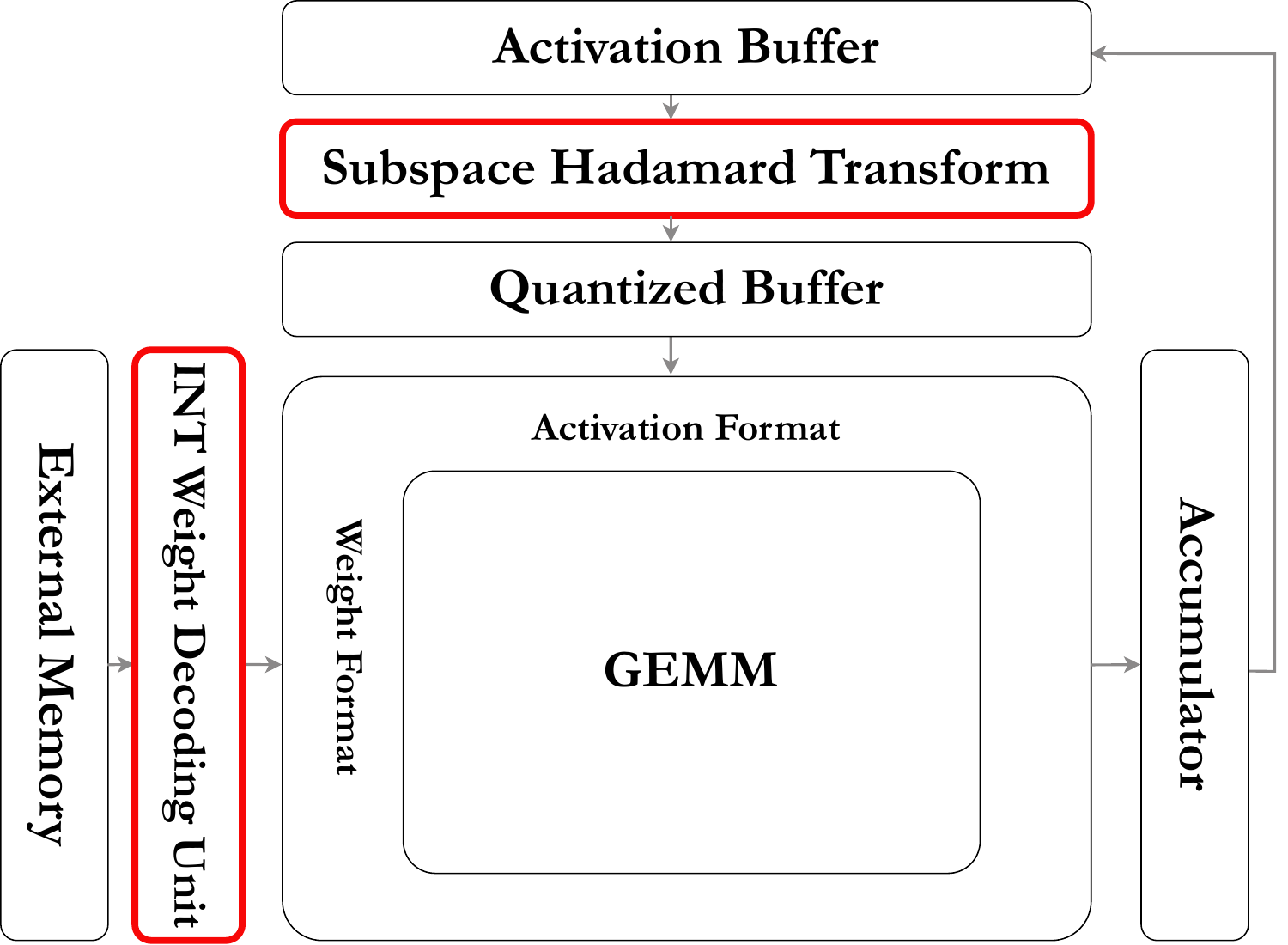}
    \caption{TQIR system overview, highlighting the INT Weight Decoding Unit corresponding to FP-Encoded INT-Decoding and activation-side Subspace Hadamard Transform Unit.}
    \label{fig:tqir_overall_hardware_architecture}
\end{figure}

Fig.~\ref{fig:tqir_overall_hardware_architecture} shows the system integration of the TQIR decoder and activation-side SHT. The compressed weight stream is expanded by the decoder, while the activation buffer supplies the SHT and quantizer. Decoder parallelism is selected to match both the GEMM consumption rate and the 256-weight TCQ sequence granularity.

For a model with $N_w$ weights and $S$ storage bits/weight, the weight traffic per generated token is $D_{\rm wt}=N_wS/8$. Under peak external-memory bandwidth $B_{\rm mem}$ and ideal utilization, the weight-bandwidth-limited token rate is
\begin{equation}
\mathrm{TPS}_{\rm BW}=\frac{8B_{\rm mem}}{N_wS}.
\label{eq:tps}
\end{equation}
For the 7B reference system with the ideal LPDDR5T-9600 bandwidth of 76.8~GB/s~\cite{edge_LPDDR5}, S4, S3, and S2 require decoded-weight rates of 153.6, 204.8, and 307.2~Gweights/s, respectively. A 256-weight/cycle decoder therefore requires clock frequencies of 0.6, 0.8, and 1.2~GHz for the three formats. The worst-case S2 point sets the 1.2-GHz implementation target. Moreover, the 256-wide tile matches both the trellis sequence length and the weight-consumption granularity of a $16\times16$ PE array~\cite{PE_array}.

For the PE array, we adopt the $16\times16$ row-wise, output-stationary dot-product array described in~\cite{PE_array} as our target array. In each cycle, the decoder supplies one 256-weight tile. The 256 weights and 16 activations are broadcast along the columns. The resulting 256 products are reduced into 16 row partial sums and accumulated locally. The trellis tile size, decoder output width, and PE consumption rate are therefore identical, eliminating the need for a width-conversion buffer between decompression and GEMM.

At the maximum 1.2-GHz point, decoder and PE rates are both
\begin{equation}
R_{\rm dec}=R_{\rm PE}=256\times1.2\times10^9=307.2~\mathrm{Gweights/s}.
\label{eq:dec_pe_rate}
\end{equation}
For a 7B model, this rate corresponds to a compute/decode ceiling of 43.89~tokens/s. The analytical token rate, in tokens per second (TPS), is therefore
\begin{equation}
\mathrm{TPS}_{\rm sys}=\min(\mathrm{TPS}_{\rm BW},\mathrm{TPS}_{\rm dec},\mathrm{TPS}_{\rm PE}).
\label{eq:system_balance}
\end{equation}
As shown in Table~\ref{tab:system_balance}, S2 fully utilizes the memory, decoder, and PE rates at 1.2~GHz. S3 and S4 remain memory-bandwidth limited, using only 67\% and 50\%, respectively, of a decoder/PE fabric operated at the worst-case clock. 
For fixed-format deployments, the corresponding 0.6-, 0.8-, and 1.2-GHz settings for S4, S3, and S2 are summarized with the implementation results in Table~\ref{tab:e2e}.

\begin{table}[t]
\centering
\caption{Decoder--PE--bandwidth balance at 256 weights/cycle and 1.2~GHz.}
\label{tab:system_balance}
\scriptsize
\setlength{\tabcolsep}{3.3pt}
\begin{tabular}{lrrrr}
\toprule
Format & BW TPS & Dec./PE TPS & System TPS & Dec./PE util. \\
\midrule
S4W8A8 & 21.94 & 43.89 & 21.94 & 50\% \\
S3W4A4 & 29.26 & 43.89 & 29.26 & 67\% \\
S2W4A4 & 43.89 & 43.89 & 43.89 & 100\% \\
\bottomrule
\end{tabular}
\end{table}

\subsection{Configurable Multi-Format Decoder}

\begin{figure}[htbp]
    \centering
    \includegraphics[width=0.95\linewidth]{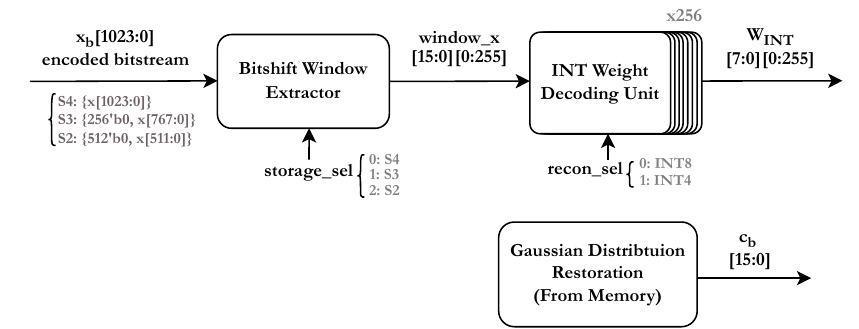}
    \caption{Architecture of the Configurable Multi-Format TQIR Weight Decoder. The storage-format-controlled window extractor supplies 256 parallel INT Weight Decoding Units, while reconstruction-format control generates INT8 or INT4 weights and the precomputed scaling factor is separately forwarded to INT GEMM.}
    \label{fig:configurable_multi_format_decoder}
\end{figure}

The decoder is derived by progressively simplifying the original 1MAD reconstruction into static integer hardware. All TQIR modes use the same 16-bit 1MAD state and the same LCG mapping; the storage precision determines only the stride between adjacent state windows, whereas the reconstruction precision determines only the final integer range. The dominant arithmetic can therefore be shared across S4W8A8, S3W4A4, and S2W4A4.

\subsubsection{INT Weight Decode Unit}
As shown in Fig.~\ref{fig:configurable_multi_format_decoder}, the first simplification is static bitshift-window extraction. For storage width $S$, the $t$th codeword input is
\begin{equation}
x_t=\mathcal{B}_{\rm ext}[tS+:16],\qquad 0\le t<256,
\label{eq:window}
\end{equation}
where $\mathcal{B}_{\rm ext}$ appends the leading $16-S$ bits to implement tail-biting. Because every offset is known at design time, all windows are generated by fixed slicing and wiring rather than iterative trellis-state recovery. The extraction stage therefore requires no arithmetic unit, codebook SRAM, or state-transition controller.

The extracted state is evaluated by the 1MAD LCG as
\begin{equation}
p=(Ax+B)\bmod 2^{32},
\label{eq:one_mad_lcg_hw}
\end{equation}
with $A=34{,}038{,}481$ and $B=76{,}625{,}530$. A direct implementation would replicate a general constant multiplier in every decoding lane. Since $A$ is fixed, it is expressed in canonical signed-digit (CSD) form as
\begin{equation}
\begin{aligned}
A={}&2^{25}+2^{19}-2^{15}-2^{13}+2^{10}\\
&-2^{8}-2^{6}+2^{4}+1,
\end{aligned}
\label{eq:csd}
\end{equation}
so $Ax$ is realized with hardwired shifts and signed additions. The modulo-$2^{32}$ operation is obtained by retaining only the lower 32 bits, avoiding an explicit modulo unit.

\begin{figure}[htbp]
    \centering
    \includegraphics[width=0.72\linewidth]{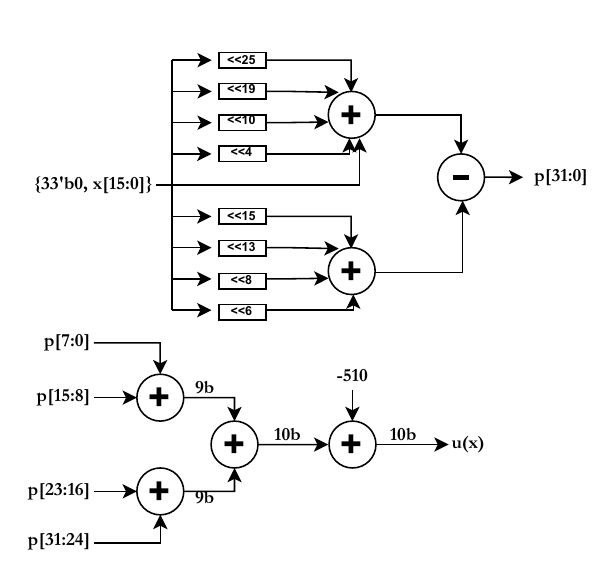}
    \caption{Datapath of the CSD-optimized INT Weight Decoding Unit. The constant LCG multiplication is replaced by fixed shift-and-add logic, followed by byte summation and centering.}
    \label{fig:modified_1mad_compute_unit}
\end{figure}

As illustrated in Fig.~\ref{fig:modified_1mad_compute_unit}, the 32-bit result is partitioned into four bytes and reduced through a balanced adder tree,
\begin{equation}
u(x)=\sum_{j=0}^{3}\mathrm{byte}_{j}(p)-510.
\label{eq:bytesum_hw}
\end{equation}
The format-dependent power-of-two shift in (\ref{eq:int_grid}) then maps the centered integer to the target INT8 or INT4 range. Thus, the online reconstruction is reduced to
\begin{equation}
\begin{split}
\text{bitstream}
&\rightarrow \text{fixed 16-bit window}
\rightarrow \text{CSD shift--add LCG}\\
&\rightarrow \text{byte sum/centering}
\rightarrow \text{INT8/INT4 range scaling},
\end{split}
\label{eq:decoder_simplification_flow}
\end{equation}
without floating-point normalization or reconstruction in the weight path. The integer output $W_{\rm INT}$ is sent directly to the GEMM array, while the shared restoration factor $s_W$ is forwarded separately for subsequent output scaling.

The unit-level simplifications are extended to the system decoder in Fig.~\ref{fig:configurable_multi_format_decoder}. A storage-format-controlled extractor supplies 256 state windows per cycle, matching one TCQ sequence and the 256-weight/cycle PE consumption rate. The 256 INT Weight Decoding Units share the same 1MAD structure for all modes, while \texttt{storage\_sel} changes only the bit-window stride and \texttt{recon\_sel} selects the INT8/INT4 range mapping. Format-dependent control is therefore confined to the input and output boundaries instead of being replicated throughout the main arithmetic datapath.

\subsection{Format-Customized Decoder}
\label{subsec:format_customized_decoder_optimization}

The configurable decoder minimizes design duplication across operating modes, but its 256 lanes still evaluate the 1MAD mapping independently. Once the storage format is fixed, neighboring bitshift windows exhibit deterministic overlap: an $S$-bit format introduces only $S$ new bits per decoded weight while reusing the remaining $16-S$ bits. The Format-Customized Decoder exploits this structure through Cross-Input Common Subexpression Elimination (CSE), moving repeated arithmetic from the lane level into shared partial-product and accumulation networks.

For S4, partition the encoded stream into four-bit symbols $C_i$. The input window of lane $t$ is
\begin{equation}
X^{(t)}=C_{t+3}+(C_{t+2}\ll4)+(C_{t+1}\ll8)+(C_t\ll12),
\label{eq:s4_overlapped_input_window}
\end{equation}
so consecutive windows reuse three of four symbols. Defining a shared low-bit partial product
\begin{equation}
P_i=(AC_i)\bmod2^{32},
\label{eq:s4_shared_partial_product}
\end{equation}
allows each $P_i$ to be generated once and reused by multiple neighboring outputs rather than repeatedly evaluating the complete $AX^{(t)}$ operation.

\begin{figure}[tbh]
    \centering
    \includegraphics[width=\linewidth]{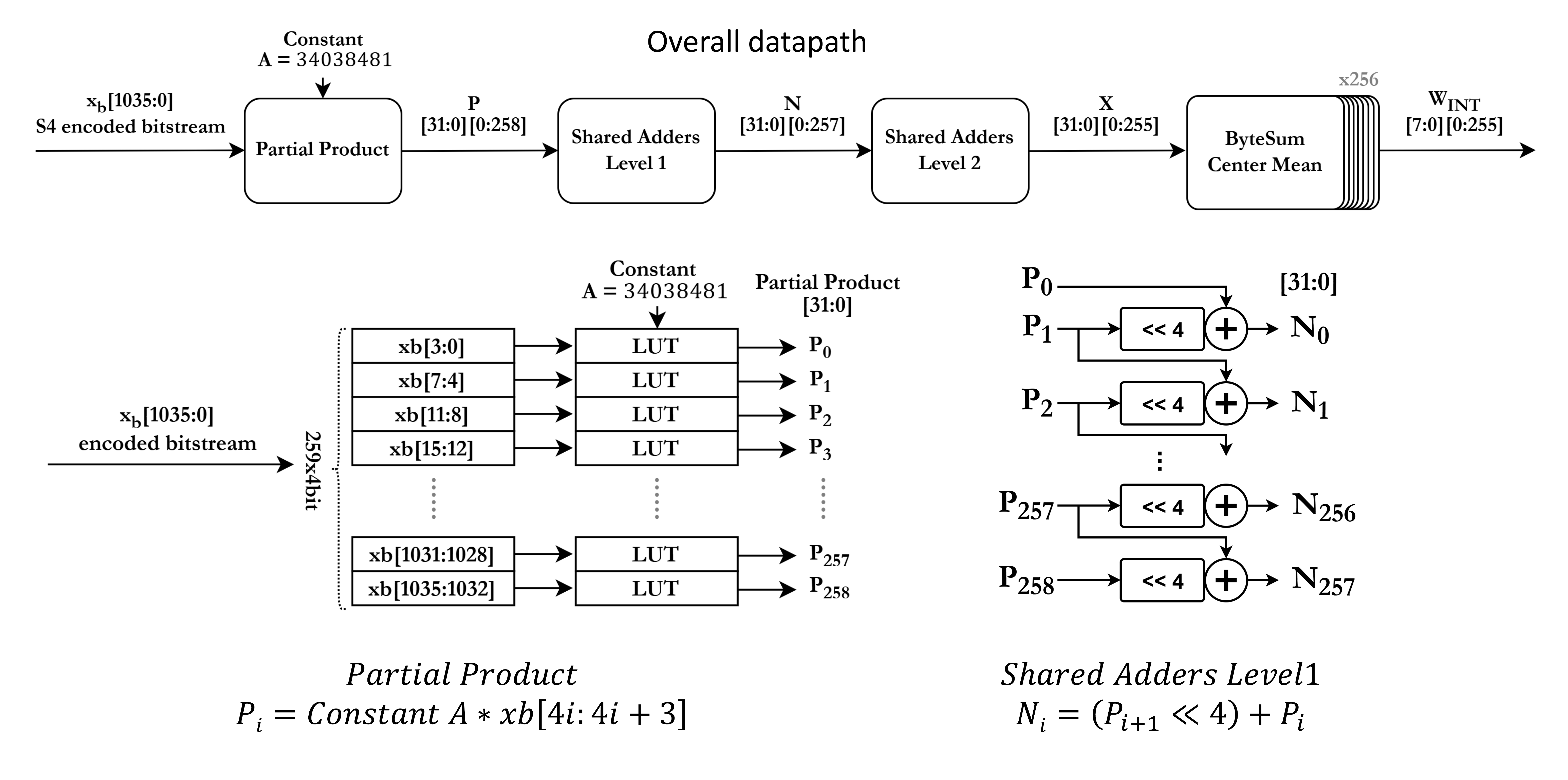}
    \caption{S4 Format-Customized Decoder using Cross-Input CSE.}
    \label{fig:s4_customized_decoder}
\end{figure}

Fig.~\ref{fig:s4_customized_decoder} shows the resulting S4 architecture. For one 256-weight tile, tail-biting extends the 1024-bit S4 stream to 1036 bits, which is partitioned into 259 four-bit symbols. Each symbol produces one $P_i$ using a small fixed mapping or equivalent shift-and-add logic, as shown in Fig.~\ref{fig:s4_customized_decoder}. The first shared-adder level combines adjacent partial products,
\begin{equation}
N_i=(P_{i+1}\ll4)+P_i,
\label{eq:s4_shared_adder_level1}
\end{equation}
and the second level reuses these overlapping eight-bit partial results,
\begin{equation}
Y_i=\left[(N_{i+2}\ll8)+N_i+B\right]\bmod2^{32}.
\label{eq:s4_shared_adder_level2}
\end{equation}
Substituting (\ref{eq:s4_shared_adder_level1}) into (\ref{eq:s4_shared_adder_level2}) reconstructs the same $(AX^{(i)}+B)\bmod2^{32}$ result as the independent INT Weight Decode Unit. Therefore, Cross-Input CSE changes only the organization of the intermediate arithmetic; the trellis code, reconstruction function, 256-weight/cycle throughput, and output interface remain unchanged.

The sharing structure is format-dependent. S4 and S2 naturally partition the 16-bit state using aligned 4-bit and 2-bit strides, respectively. In contrast, the 3-bit S3 stride is not aligned with the byte boundaries of the subsequent 1MAD reduction and requires phase-dependent partial-product paths. Conceptually, the configurable design performs \emph{intra-input simplification} by replacing each lane's general arithmetic with fixed wiring and constant operations, whereas the customized design additionally performs \emph{inter-input simplification} by reusing low-bit terms across neighboring lanes. The synthesis results in Section~\ref{sec:hardware_efficiency} quantify the resulting area--energy trade-off.

\subsection{Scheduling-Aware \texorpdfstring{$H_{128}$}{H128} Unit}
An $H_{128}$ FWHT contains seven stages with 64 butterfly pairs per stage, for a total of 448 pair operations per transform. 
Increasing pair parallelism replicates the butterfly resources and therefore increases SHT area and power. Provisioning the
transform for the instantaneous PE input bandwidth is therefore unnecessary when the transform can instead be scheduled
against its consumption deadline.
Instead, \tqir schedules the transform according to the production and consumption deadlines of 128-element activation groups. A four-pair unit then completes one group in 112 cycles. 

\begin{figure}[tbh]
    \centering
    \includegraphics[width=0.90\linewidth]{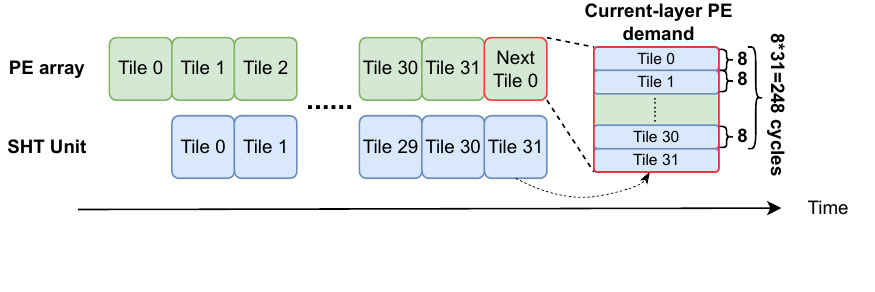}
    \caption{Hiding the $H_{128}$ SHT latency behind previous PE-array execution.}
    \label{fig:h128_layer_level_scheduling}
\end{figure}

The scheduling opportunity follows from the output-stationary production pattern. For a $4096\times4096$ linear layer, a 16-output block becomes final after $4096/16=256$ accumulation cycles, so a complete 128-element activation group is produced every $8\times256=2048$ cycles. The SHT unit begins processing when a group becomes available and writes the transformed result to a small buffer while the previous layer continues to execute. As illustrated in Fig.~\ref{fig:h128_layer_level_scheduling}, the tighter final-group constraint provides a $31\times8=248$-cycle preparation window before the next-layer group is consumed. The four-pair SHT completes the group in 112 cycles, leaving a 136-cycle scheduling margin. Thus, under the evaluated schedule, the SHT can be kept off the PE critical path.

\section{Experimental Evaluation}
\subsection{Setup}
The algorithmic evaluation uses LLaMA-2-7B and LLaMA-3-family models~\cite{llama2,llama3}, with WikiText2 and C4 perplexity as the primary metrics. The 1MAD state length is 16 bits, each TCQ sequence contains 256 weights, and $H_{128}$ is the default subspace configuration. For 8-bit activation quantization (A8), we use symmetric per-channel quantization, whereas A4 activations use symmetric group-wise quantization with a group size of 32. The configurable decoder is synthesized in a TSMC 16-nm process using Synopsys Design Compiler, with power estimated using PrimeTime PX.

\subsection{Hardware Efficiency}
\label{sec:hardware_efficiency}

\subsubsection{Comparisons}
Table~\ref{tab:hardware} compares the full configurable TQIR decoder with Ecco~\cite{expresult_entropybasedmethod_ecco} and Codeword Decomposition~\cite{entro_codeworddecompos}. TQIR avoids serial symbol parsing through its fixed-rate stream and a large explicit codebook through computed-code reconstruction. For cross-process comparison, silicon area is normalized to 16~nm using
\begin{equation}
A_{16}=A_{\rm reported}\left(\frac{16}{N_{\rm process}}\right)^2.
\label{eq:area_norm}
\end{equation}
Power is not technology-normalized. 
TQIR achieves 6034.5~GB/s/mm$^2$ of decoded-output bandwidth per normalized area, 13.93$\times$ and 1.52$\times$ those of Ecco and Codeword Decomposition, respectively. These ratios characterize decoder-area efficiency rather than complete-accelerator efficiency.

\begin{table}[t]
\centering
\caption{Full-decoder hardware comparison. Area is normalized to 16~nm and power is not normalized.
}
\label{tab:hardware}
\scriptsize
\setlength{\tabcolsep}{3.0pt}
\begin{tabular}{lccc}
\toprule
Metric & TQIR & Ecco~\cite{expresult_entropybasedmethod_ecco} & Codeword~\cite{entro_codeworddecompos} \\
\midrule
Process (nm) & 16 & 7 & 40 \\
Frequency (GHz) & 1.20 & 1.41 & 1.50 \\
Area (mm$^2$) & 0.0509 & 3.190 & 0.0024 \\
Power (mW) & 53.0 & 4820 & 1.14 \\
Normalized area (mm$^2$) & 0.0509 & 16.666 & 0.00038 \\
Output weights/cycle & 256 & 2560 & 1 \\
Decoded output (GB/s) & 307.2 & 7219.2 & 1.5 \\
Area eff. (GB/s/mm$^2$) & \textbf{6034.5} & 433.2 & 3978.1 \\
\bottomrule
\end{tabular}
\end{table}

\subsubsection{Configurable and Customized Decoders}
Table~\ref{tab:config_custom} compares the configurable and customized decoders at their format-matched frequencies. Decoder energy is calculated as $E_{\mathrm{dec}}=P/T_{\mathrm{weight}}$. 
Cross-Input CSE improves area efficiency by 61.2\%, 8.0\%, and 2.2\% for S4, S3, and S2, respectively, with S4 benefiting most from format-specific computation sharing. However, the customized implementations consume higher energy per decoded weight: 0.324 versus 0.183~pJ/weight for S4, 0.311 versus 0.193~pJ/weight for S3, and 0.396 versus 0.173~pJ/weight for S2. Cross-Input CSE is therefore primarily an area-efficiency optimization for fixed-format deployment, whereas the configurable decoder trades area for multi-format flexibility.

\begin{table}[t]
\centering
\caption{Configurable versus customized decoders at format-matched frequency. 
}
\label{tab:config_custom}
\resizebox{\columnwidth}{!}{%
\begin{tabular}{llrrrr}
\toprule
Format & Design & Area ($\mu$m$^2$) & Power (mW) &
pJ/weight & Area-eff. gain \\
\midrule
\multirow{2}{*}{S4W8A8}
& Config. & 44141 & 28.1 & 0.183 & -- \\
& Custom. & 27383 & 49.8 & 0.324 & 61.2\% \\
\midrule
\multirow{2}{*}{S3W4A4}
& Config. & 47996 & 39.6 & 0.193 & -- \\
& Custom. & 44454 & 63.6 & 0.311 & 8.0\% \\
\midrule
\multirow{2}{*}{S2W4A4}
& Config. & 50949 & 53.0 & 0.173 & -- \\
& Custom. & 49812 & 121.8 & 0.396 & 2.2\% \\
\bottomrule
\end{tabular}}
\end{table}

\subsubsection{Ideal Throughput Evaluation}
Using the same 7B/76.8-GB/s ideal-bandwidth model as Table~\ref{tab:system_balance}, Table~\ref{tab:e2e} reports \emph{analytical weight-bandwidth ceilings} rather than measured chip-level throughput. KV-cache, activation, control, and other system traffic are excluded to isolate the effect of trellis storage width. The decoder rows use the configurable implementation at each format-matched clock. The shared four-pair $H_{128}$ unit adds only 0.00127~mm$^2$ and 2.33~mW and is omitted from these decoder columns; its 112-cycle latency fits within the 248-cycle preparation window of Fig.~\ref{fig:h128_layer_level_scheduling}.

\begin{table}[t]
\centering
\caption{Throughput (tokens per second: TPS) for LLaMA-2-7B. Token rates are ideal weight-bandwidth ceilings at 76.8~GB/s; quality uses sequence length 2048.}
\label{tab:e2e}
\scriptsize
\setlength{\tabcolsep}{3.1pt}
\textit{(a) Model representation and bandwidth}\\
\begin{tabular}{lrrrr}
\toprule
Mode & $S/W/A$ & GB/token & Ideal TPS & WikiText2/C4 \\
\midrule
FP16 & 16/16/16 & 14.000 & 5.49 & 5.47/6.97 \\
TQIR quality & 4/8/8 & 3.500 & 21.94 & 5.53/7.04 \\
TQIR balanced & 3/4/4 & 2.625 & 29.26 & 5.81/7.34 \\
TQIR bandwidth & 2/4/4 & 1.750 & 43.89 & 6.60/8.22 \\
\bottomrule
\end{tabular}

\vspace{2pt}
\textit{(b) Corresponding configurable decoder settings for (a) condition}\\
\begin{tabular}{lrrrr}
\toprule
Mode & Rate (Gwt/s) & Clock (GHz) & Area (mm$^2$) & Power (mW) \\
\midrule
S4W8A8 & 153.6 & 0.60 & 0.0441 & 28.1 \\
S3W4A4 & 204.8 & 0.80 & 0.0480 & 39.6 \\
S2W4A4 & 307.2 & 1.20 & 0.0509 & 53.0 \\
\bottomrule
\end{tabular}
\end{table}

\subsection{Algorithmic Ablations}
\subsubsection{Integer-Reproducible Trellis Decoding}

Table~\ref{tab:fpencodeINTdecode_result} isolates decoded-weight precision with FP16 activations. 
For the deployment-selected formats, S4 INT8 exactly matches the S4 FP16-reconstruction WikiText2 perplexity (5.16), while S3 and S2 INT4 differ from their FP16-reconstruction counterparts by only 0.01 (5.29 versus 5.30 and 6.02 versus 6.03, respectively). These controlled results directly support FP-Encoded INT-Decoding: the target integer grid can be incorporated into offline trellis selection without materially changing the selected-path quality.

\begin{table}[t]
\centering
\caption{Decoded-weight precision ablation on LLaMA-2-7B (sequence length 4096). The experimental setup uses the QTIP setup (global Hadamard transform) for fair comparison.}
\label{tab:fpencodeINTdecode_result}
\footnotesize
\setlength{\tabcolsep}{4.5pt}
\begin{tabular}{ccccc}
\toprule
$S$ & $W$ & WikiText2 & C4 & BoolQ \\
\midrule
FP16 & FP16 & 5.12 & 6.63 & 77.74\% \\
\midrule
4 & FP16 & 5.16 & 6.68 & 76.88\% \\
4 & INT8 & 5.16 & 6.68 & 77.76\% \\
4 & INT4 & 5.28 & 6.87 & 75.10\% \\
\midrule
3 & FP16 & 5.29 & 6.86 & 76.17\% \\
3 & INT8 & 5.28 & 6.86 & 77.98\% \\
3 & INT4 & 5.30 & 6.89 & 75.35\% \\
\midrule
2 & FP16 & 6.02 & 7.81 & 72.62\% \\
2 & INT8 & 6.01 & 7.82 & 74.06\% \\
2 & INT4 & 6.03 & 7.81 & 74.52\% \\
\bottomrule
\end{tabular}
\end{table}

The same weight-only experiment also reports downstream zero-shot accuracy. Table~\ref{tab:weight_only_zero_shot} shows that the integer-reconstructed paths retain the overall trend of the FP16 baseline across ARC-Challenge, ARC-Easy, BoolQ, and WinoGrande. As expected, the two-bit storage point produces the largest degradation; the purpose of this table is to expose this trade-off rather than to imply a uniform accuracy gain based on perplexity alone.

\begin{table}[t]
\centering
\caption{Weight-only zero-shot accuracy on LLaMA-2-7B for representative TQIR reconstruction points.}
\label{tab:weight_only_zero_shot}
\footnotesize
\setlength{\tabcolsep}{3.0pt}
\begin{tabular}{lcccc}
\toprule
Configuration & ARC-C & ARC-E & BoolQ & WinoGrande \\
\midrule
FP16 baseline & 43.43 & 76.26 & 77.74 & 68.98 \\
$S_4W_8$ & 42.66 & 76.38 & 77.76 & 68.27 \\
$S_3W_4$ & 41.97 & 75.42 & 75.35 & 69.21 \\
$S_2W_4$ & 37.88 & 72.55 & 74.52 & 66.29 \\
\bottomrule
\end{tabular}
\end{table}

For an additional weight-only comparison, Table~\ref{tab:weight_only_prior_context} compares TQIR with other quantization methods. TQIR here uses FP-Encoded INT-Decoding and Subspace-Aligned Trellis Encoding. At four bits, TQIR $S_4W_8$ closely matches QTIP while replacing FP reconstruction with INT8. At three bits, $S_3W_4$ remains near QTIP and the vector-quantization baselines. At two bits, TQIR incurs a modest penalty relative to QTIP but remains far more stable than scalar GPTQ in the reported experiment.

\begin{table}[tbh]
\centering
\caption{Comparison of weight-only quantization methods on LLaMA-2-7B using WikiText2.}
\label{tab:weight_only_prior_context}
\footnotesize
\setlength{\tabcolsep}{3.6pt}
\begin{tabular}{lccc}
\toprule
Method & 4-bit & 3-bit & 2-bit \\
\midrule
GPTQ~\cite{expresult_gptq} & 5.49 & 8.06 & 50.75 \\
GPTVQ~\cite{gptvq} & 5.27 & 5.44 & 6.71 \\
QuIP\#~\cite{quip2} & 5.19 & 5.41 & 6.19 \\
AQLM~\cite{aqlm} & 5.21 & 5.46 & 6.64 \\
VPTQ~\cite{methodvq_vptq} & 5.26 & 5.43 & 6.13 \\
QTIP~\cite{qtip} & 5.17 & \textbf{5.28} & \textbf{5.86} \\
TQIR & \textbf{5.16} & 5.30 & 6.01 \\
\bottomrule
\end{tabular}
\end{table}
\subsubsection{Hadamard Placement}
\begin{table}[tbh]
\centering
\caption{Reverse-Hadamard placement ablation for $S_2W_4A_4$.
Lower perplexity is better.}
\label{fig:overall_reversehada_ppl}
\footnotesize
\setlength{\tabcolsep}{4.2pt}
\renewcommand{\arraystretch}{1.08}
\begin{tabular}{llcc}
\toprule
Activation Quant. & Hadamard Placement & WikiText2 & C4 \\
\midrule
\multirow{2}{*}{Per-channel}
& Activation-side & \textbf{6.484} & \textbf{8.406} \\
& Weight-side     & 2425 & 2473 \\
\midrule
\multirow{2}{*}{Group-wise (32)}
& Activation-side & \textbf{6.195} & \textbf{8.036} \\
& Weight-side     & 6.270 & 8.156 \\
\bottomrule
\end{tabular}
\end{table}

\begin{table}[tbh]
\centering
\caption{
Subspace-Hadamard block-size ablation for $S_2W_4A_4$ using the original $16\times16$ TCQ tile. Computation reduction is measured relative to the global Hadamard transform.
}
\label{tab:Subspace_Hadamard_pplresults}
\scriptsize
\setlength{\tabcolsep}{2.8pt}
\renewcommand{\arraystretch}{1.05}
\begin{tabular}{lcccc}
\toprule
Transform & WikiText2 & C4 & FWHT Stages & \shortstack{Computation\\Reduction} \\
\midrule
Global & 6.20 & 8.04 & $\log_2N$ & 0.0\% \\
$H_{256}$ & 6.13 & 8.03 & 8 & 42.2\% \\
$H_{128}$ & \textbf{6.20} & \textbf{8.04} & 7 & \textbf{49.5\%} \\
$H_{64}$ & 6.30 & 8.18 & 6 & 56.7\% \\
$H_{32}$ & 7.41 & 10.00 & 5 & 63.9\% \\
$H_{16}$ & 2933 & 2397 & 4 & 71.1\% \\
\bottomrule
\end{tabular}
\end{table}

\begin{table}[tbh]
\centering
\caption{Subspace-aligned encoding ablation for $S_2W_4A_4$.}
\label{tab:subspace_block_encode_tile}
\footnotesize
\setlength{\tabcolsep}{4pt}
\begin{tabular}{cccc}
\toprule
SHT & Tile & WikiText2 & C4 \\
\midrule
$H_{256}$ & $16\times16$ & 6.131 & 8.028 \\
$H_{256}$ & $256\times1$ & 6.117 & 8.013 \\
$H_{128}$ & $16\times16$ & 6.196 & 8.041 \\
$H_{128}$ & $128\times2$ & \textbf{6.134} & \textbf{8.013} \\
$H_{64}$ & $16\times16$ & 6.303 & 8.181 \\
$H_{64}$ & $64\times4$ & 6.206 & 8.066 \\
$H_{32}$ & $16\times16$ & 7.412 & 10.002 \\
$H_{32}$ & $32\times8$ & 7.340 & 9.755 \\
\bottomrule
\end{tabular}
\end{table}

\begin{table}[tbh]
\centering
\caption{Representative INT8/INT4 GEMM deployment results. Baselines are compared only within each matched experiment group.}
\label{tab:deployment_quality}
\footnotesize
\setlength{\tabcolsep}{3.8pt}
\begin{tabular}{lccc}
\toprule
Model / group & Configuration & Storage & WikiText2 \\
\midrule
LLaMA-2-7B W8A8 & FP16 baseline & 16 b & 5.47 \\
 & $S_4W_8A_8$ & 4 b & 5.53 \\
\midrule
LLaMA-2-7B W4A4 & FP16 baseline & 16 b & 5.47 \\
 & $S_3W_4A_4$ & 3 b & 5.81 \\
 & $S_2W_4A_4$ & 2 b & 6.60 \\
\midrule
LLaMA-3-8B W8A8 & FP16 baseline & 16 b & 6.14 \\
 & $S_4W_8A_8$ & 4 b & 6.27 \\
\midrule
LLaMA-3-8B W4A4 & FP16 baseline & 16 b & 6.14 \\
 & $S_3W_4A_4$ & 3 b & 7.02 \\
 & $S_2W_4A_4$ & 2 b & 8.73 \\
\bottomrule
\end{tabular}
\end{table}

\begin{table}[tbh]
\centering
\caption{Cross-model perplexity of TQIR deployment configurations.}
\label{tab:cross_model_tqir}
\scriptsize
\setlength{\tabcolsep}{2.8pt}
\begin{tabular}{lccc}
\toprule
Model & Configuration & WikiText2 & C4 \\
\midrule
\multirow{4}{*}{LLaMA-3.1-8B}
& FP16 & 6.24 & 8.96 \\
& $S_4W_8A_8$ & 6.37 & 9.11 \\
& $S_3W_4A_4$ & 7.06 & 10.11 \\
& $S_2W_4A_4$ & 8.68 & 12.05 \\
\midrule
\multirow{4}{*}{LLaMA-3.1-8B-Inst.}
& FP16 & 7.22 & 10.39 \\
& $S_4W_8A_8$ & 7.34 & 10.58 \\
& $S_3W_4A_4$ & 8.02 & 11.55 \\
& $S_2W_4A_4$ & 9.76 & 13.81 \\
\midrule
\multirow{4}{*}{LLaMA-3.2-3B}
& FP16 & 7.82 & 10.44 \\
& $S_4W_8A_8$ & 7.96 & 10.65 \\
& $S_3W_4A_4$ & 8.74 & 11.85 \\
& $S_2W_4A_4$ & 11.09 & 14.56 \\
\bottomrule
\end{tabular}
\end{table}

Table~\ref{fig:overall_reversehada_ppl} evaluates reverse-Hadamard placement under $S_2W_4A_4$. With per-channel activation quantization,
weight-side transformation causes severe perplexity degradation because the original activation outliers are quantized before Hadamard smoothing. Relocating the transform to the activation side reduces WikiText2 and C4 perplexity from 2425 and 2473 to 6.484 and 8.406, respectively. With 32-element group-wise quantization, local scaling substantially reduces the sensitivity to transform placement; activation-side and weight-side transformation achieve 6.195 versus 6.270 on WikiText2 and 8.036 versus 8.156 on C4. 
These results support activation-side relocation because it preserves the integer weight path and yields lower perplexity
under both evaluated activation-quantization granularities.

\subsubsection{Subspace Size and Trellis Alignment}
Table~\ref{tab:Subspace_Hadamard_pplresults} quantifies the quality--computation trade-off of the Subspace Hadamard Transform. Very small subspaces provide the largest computation reduction but, consistent with insufficient cross-channel mixing, cause severe perplexity degradation: $H_{16}$ and $H_{32}$ reduce FWHT computation by 71.1\% and 63.9\%, respectively. $H_{64}$ still saves 56.7\% of the transform computation, but its WikiText2/C4 perplexity increases to 6.30/8.18. In contrast, $H_{128}$ reduces the transform computation by 49.5\% while matching the evaluated global-transform perplexity of 6.20/8.04. Increasing the block size to $H_{256}$ lowers the computation reduction to 42.2\% and provides only a marginal perplexity improvement to 6.13/8.03. We therefore select $H_{128}$ as the quality--complexity knee point: it preserves global-transform model quality while removing approximately half of the runtime Hadamard computation.

Table~\ref{tab:subspace_block_encode_tile} then shows the effect of the proposed tile reshaping. For $H_{128}$, changing the QTIP $16\times16$ tile to the aligned $128\times2$ tile reduces WikiText2 from 6.196 to 6.134 and C4 from 8.041 to 8.013. The aligned tile consistently lowers perplexity across $H_{32}$--$H_{256}$ without changing the number of encoded weights, trellis dimension, or storage width. 

\subsection{Deployment Model Quality}
Table~\ref{tab:deployment_quality} summarizes representative integer-GEMM deployment points. $S_4W_8A_8$ uses half the weight-fetch bits of a conventional W8A8 model while remaining close to its FP16 baseline. $S_3W_4A_4$ reduces storage by 25\% relative to four-bit weights and provides the balanced operating point; $S_2W_4A_4$ halves W4 weight traffic at a larger quality penalty.

Table~\ref{tab:cross_model_tqir} verifies the same quality--bandwidth ordering on additional base and instruction-tuned LLaMA variants. Across the evaluated models, $S_4W_8A_8$ consistently introduces the smallest perplexity increase, $S_3W_4A_4$ provides a moderate penalty at three stored bits/weight, and $S_2W_4A_4$ provides the lowest traffic at the largest penalty. The LLaMA-3.1-8B-Instruct result further indicates that the ordering is retained after instruction tuning rather than being limited to base checkpoints.

\section{Conclusion}
This paper presented \tqir, an algorithm--hardware co-design that converts ultra-low-bit trellis-compressed LLM weights into a bandwidth-matched integer-GEMM datapath. FP-Encoded INT-Decoding preserves floating-point-domain trellis selection while generating INT8/INT4 weights at runtime. Activation-side transform relocation and $H_{128}$ subspaces maintain the integer weight interface while reducing transform work, and subspace-aligned tiles improve quantization quality without expanding the bitstream. The 16-nm configurable decoder sustains 307.2~Gweights/s, format customization provides format-dependent area savings, and scheduling-aware SHT folding fits within the evaluated PE schedule. Together, the model-quality, bandwidth, and implementation results support the system-level potential of trellis compression when storage precision, runtime numerical format, decompression architecture, and transform scheduling are co-designed under the stated system assumptions.

\bibliographystyle{IEEEtran}
\bibliography{references}

\end{document}